\documentclass[12pt]{article}

\usepackage[francais,english]{babel}
\usepackage{amssymb}
\usepackage{latexsym}
\usepackage{graphicx}
\usepackage[latin1]{inputenc} 
\usepackage{color}
\usepackage{multirow}
\usepackage{lineno}
\usepackage{url}

\begin{document}

\baselineskip 0.8cm

\thispagestyle{empty}

{\Large {\bf Non-linear regression models for Approximate Bayesian Computation}}\par

\vspace{0.5cm}



\noindent  Michael GB Blum$\star$,\par
\noindent  Olivier Fran\c cois$\dag$

\vspace{1cm}

\noindent $\star$Centre National de la Recherche Scientifique, $\dag$Institut National Polytechnique de Grenoble, TIMC-IMAG, Faculty of Medicine of Grenoble, 38706 La Tronche France.

\noindent Corresponding author: Michael Blum. \par
\noindent Address: TIMC-IMAG, Facult\'e de  M\'edecine de Grenoble, 38706 La Tronche, France.\par
\noindent Tel: +33 456 520 065. Fax: +33 456 520 055. \par
\noindent E.mail: {\tt michael.blum@imag.fr}\par

\vspace{0.5cm}


\newpage
\begin{center}
{\bf Abstract}
\end{center}

\vspace{1cm}

Approximate Bayesian inference on the basis of summary statistics is well-suited
to complex problems for which the likelihood is either mathematically or
computationally intractable. However the methods that use rejection suffer from the
curse of dimensionality when the number of summary statistics is increased. Here we propose a machine-learning approach to the estimation of the posterior density by introducing two innovations. The new method fits a nonlinear
conditional heteroscedastic regression of the parameter on the summary
statistics, and then adaptively improves estimation using importance sampling. The new algorithm is compared to the state-of-the-art approximate Bayesian methods, and achieves considerable reduction of the computational burden in two examples of inference in statistical genetics and in a queueing model.

\vspace{.5cm}


\vspace{.5cm}

\newpage

\section{Introduction}

Making use of simulations to perform non-Bayesian inference in models for which the likelihood is neither analytically solvable
 nor computationally tractable has a well-established methodology that finds its roots at least in the seminal papers of Diggle
 and Gratton (1984) and Gourieroux et al. (1993). This approach bypasses explicit likelihood functions by simulating from an {\it implicit statistical model} -- that is, a model defined in term of a stochastic generating mechanism.

In the Bayesian setting, there has been a growing interest in implicit statistical models for demographic inference in population genetics (Marjoram and Tavar\'e 2006). Statistical inference with population-genetic data usually requires prior knowledge on genealogical trees. As the tree is usually considered as a nuisance parameter, Bayesian Monte Carlo is a natural approach to average over high-dimensional tree space.  Although many likelihood-based methods have been proposed in this framework, these methods are restricted to particular demographic and genetic processes (Stephens and Donnelly 2000; Wilson et al. 2003; Kuhner 2006; Hey and Nielsen 2007). Alternatively, likelihood-free methods, named approximate Bayesian computation (ABC) after Beaumont et al. (2002), have gained many advocates in the recent years. The principle of standard ABC (Tavar\'e et al. 1997; Pritchard et al. 1999; Beaumont et al. 2002) relies on the simulation of large numbers of data sets using parameters drawn from the prior distribution. A set of summary statistics is then calculated for each simulated sample, and compared with the values for the observed sample.  Parameters that have generated summary statistics close enough to the observed data are retained to form an approximate sample from the posterior distribution.

Approximate Bayesian estimation algorithms -- that were originally all based on rejection algorithms -- can be classified into three broad categories, resembling the mainstream methods that are applied in standard computational Bayesian statistics (Gelman et al. 2003). The first class of methods relies on the direct rejection algorithm as described in the previous paragraph (Tavar\'e et al. 1997; Pritchard et al. 1999). The second class of algorithms mimics Markov chain Monte Carlo methods (MCMC, Robert and Casella 2004), embedding simulations from the implicit model in the updating step of the stochastic algorithm (Marjoram et al. 2003;  Bortot et al. 2007). The MCMC-ABC algorithm takes into account the distance between the simulated and the observed summary statistics into the Metropolis-Hastings acceptance rule. The third class of algorithms shares similarity with the recently introduced sequential Monte Carlo samplers (SMC, Liu 2001). The main SMC-ABC algorithm combines ideas underlying rejection methods and sequential importance sampling (Sisson et al. 2007; Beaumont et al. 2008; Toni et al. 2009).

Nevertheless, a severe limitation of rejection-based generative algorithms arises when the dimensionality of the set of summary statistics increases. Because the three classes of methods attempt to sample from a small multidimensional sphere around the observed summary statistics, all of them suffer from the curse of dimensionality (see e.g. H\"ardle et al. 2004). To overcome this problem, Beaumont et al. (2002) allowed larger acceptance rates in the rejection algorithm, ranging up to $20$ percent of the simulated values, and then performed local linear adjustment in order to correct for the discrepancy between the simulated and the observed statistics. Here, we address the curse of dimensionality issue by adopting a machine learning perspective constructing a functional relationship between the generated set of summary statistics and the model parameters. Assuming perfect construction, this relationship could ideally be utilized to produce samples from the posterior distribution by exploiting information not restricted to a subset of generated values, but to the extended set.

In the first stage, our approach infers the functional relationship linking the summary statistics to the model parameters by considering a flexible nonlinear conditional heteroscedastic (NCH) model. Flexible regression models, like neural networks, are exploited to reduce dimension and to better account for the correlation within the set of summary statistics. In the second stage, we introduce an adaptive version of the NCH (ANCH) algorithm using importance sampling. The rationale of the adaptive algorithm is to iteratively limit the discrepancy between the sampling distribution and the posterior distribution, which may be particularly useful when the prior distribution is vague. In two historical examples of demographic inference in population genetics and in an example of a queueing process, we provide evidence that the NCH and the ANCH algorithms reduce the computational burden when compared to ABC with rejection and to ABC with local linear adjustment.

\section{Method}

\paragraph{Rejection and weighting.} In ABC, we assume that there is a multidimensional parameter of interest $\phi$, and the observed value {\bf s} of a set of summary statistics {\bf S} is calculated for the data. To make statistical inference, a rejection-sampling method generates random draws $(\phi_i, \textrm{\bf s}_i)$ where $\phi_i$ is sampled from the prior distribution, and $\textrm{\bf s}_i$ is measured from synthetic data, simulated from a generative model with parameter $\phi_i$.  Setting the tolerance error to a value $\delta$ and denoting by $\| . \|$ the Euclidean norm, only parameters $\phi_i$ such that $\| \textrm{\bf s} -  \textrm{\bf s}_i \| \leq \delta$ are retained. Because the summary statistics may span different scales, norms that use rescaled  distances are often considered in place of the Euclidean distance.  In our application of ABC, we rescale distances by the median absolute deviation of the simulated summary statistics $\textrm{\bf s}_i$, $i=1,\dots,M$. For this basic rejection algorithm, the accepted $\phi_i$ form a random sample from the approximate posterior distribution defined as
$$
p_\delta(\phi|\textrm{\bf s}) \propto \textrm{Pr}({\| \textrm{\bf s} -  \textrm{\bf s}_i \| \leq  \delta | \phi}) \, p(\phi) 
$$
where $p(\phi)$ denotes the prior distribution. Compared to the exact expression of the posterior distribution, the likelihood is replaced by 
$$
p(\textrm{\bf s}|\phi) \approx \textrm{Pr}({\| \textrm{\bf s} -  \textrm{\bf s}_i \| \leq  \delta | \phi}) \, .
$$
If the summary statistics are sufficient with respect to the parameter $\phi$, the approximate posterior distribution converges to the true posterior distribution as $\delta$ goes to 0. In addition, the approximate posterior distribution corresponds to the prior distribution when $\delta$ is large. Note that more generic interpretations of the ABC algorithm can be found in Wilkinson (2008).

Beaumont et al. (2002) introduced a first improvement of the standard rejection ABC algorithm in which the parameters $\phi_i$ were weighted by the values $K_{\delta}(\|\textrm{\bf s}_i - \textrm{\bf s}\|)$, where $K_{\delta}$ is the Epanechnikov kernel. Using this weighting scheme, an estimator of the posterior mean was then given by $\sum_i K_{\delta}(\|\textrm{\bf s}_i - \textrm{\bf s} \| ) \phi_i /\sum_i K_{\delta}(\|\textrm{\bf s}_i - \textrm{\bf s}\|)$. Although this was not originally stated, it can be seen that it corresponds to the Nadaraya-Watson estimator, a classic approach to nonparametric regression (Nadaraya, 1964; Watson 1964). Note that the Nadaraya-Watson smoother is known to be subject to the curse of dimensionality since the rate of convergence of the estimator decreases dramatically as the dimension of the summary statistics increases (see e.g. H\"ardle et al. 2004).

\paragraph{A local linear model.}

To avoid the curse of dimensionality, Beaumont et al. (2002) also described the posterior density as a homoscedastic linear regression model (in fact, a local-linear model) of the form
\begin{equation}
\label{eqn:locl}
\phi_i = \alpha + (\textrm{\bf s}_i  - \textrm{\bf s})^T \beta + \zeta_i, \quad i = 1, \dots, M,
\end{equation}
where $\alpha$ is an intercept, $\beta$ is a vector of regression coefficients, and the $\zeta_i$'s are independent random variates with mean zero and common variance. We further refer to this algorithm as the LocL ABC model. In the LocL model, the parameters $(\alpha,\beta)$ are inferred by minimizing the weighted least-squares criterion 

\begin{equation}
\label{eqn:leastsquare}
\sum_{i=1}^n\{\phi_i-({\bf \alpha}+({\bf s}_i -{\bf s}_{obs})^t{\bf \beta})\}^2 K_{\delta}(\|\textrm{\bf s}_i - \textrm{\bf s}\|).
\end{equation}
If equation (\ref{eqn:locl}) exactly describes the relationship between $\phi$ and ${\bf s}$, random draws of the posterior distribution can simply be obtained as $\alpha+\zeta_i$, for $i =$ $1,$ $\dots,$ $M$. Using the empirical residuals in place of the $\zeta_i$'s, the parameters are adjusted as
\begin{equation}
\label{eqn:adj_locl}
\phi_i^* =  \phi_i - (\textrm{\bf s}_i  - \textrm{\bf s})^T \hat{\beta}.
\end{equation}
Weighted by $K_{\delta}(\|\textrm{\bf s}_i - \textrm{\bf s}\|)$, the $\phi_i^*$'s, $i=1,\dots,M $, provide an approximate sample from the posterior distribution. In this approach, the choice of $\delta$ involves a bias-variance trade-off: Increasing $\delta$ reduces variance thanks to a larger sample size for fitting the regression, but also increases bias arising from departures from linearity and homoscedasticity.

\paragraph{A nonlinear conditional heteroscedastic model.} In this study, we introduce an important modification on the previously described adjustment-based ABC method for conditional density estimation. In order to minimize departures from linearity and homoscedasticity, we propose to model both the location and the scale of the response parameter, $\phi_i$, in equation (\ref{eqn:locl}) (see e.g. Fan and Yao 1998). The new regression model takes the form of a nonlinear conditional heteroscedastic model
\begin{equation}
\label{eqn:nch}
\phi_i = m(\textrm{\bf s}_i )  + \sigma(\textrm{\bf s}_i ) \times \zeta_i, \quad i = 1, \dots, M,
\end{equation}
where $m(\textrm{\bf s}_i )$ denotes the conditional expectation, $\textrm{E} [\phi | \textrm {\bf S}= \textrm{\bf s}_i]$, and $\sigma^2(\textrm{\bf s}_i )$ denotes the conditional variance, $\textrm{Var} [\phi | \textrm {\bf S}= \textrm{\bf s}_i]$.

The conditional expectation can be estimated as $\hat{m}(\textrm{\bf s}_i )$ by fitting a flexible non-linear regression model. 
The variance term is then estimated using a second regression model for the log-residuals 
\begin{equation}
\label{eqn:reg_var}
\log(\phi_i - \hat{m}(\textrm{\bf s}_i ))^2 = \log \sigma^2(\textrm{\bf s}_i ) + \xi_i, \quad i = 1, \dots, M,
\end{equation}
where the $\xi_i$'s are independent random variates with mean zero and common variance. In our forthcoming examples, we consider feed-forward neural network (FFNN) regression models (Ripley 1996; Bishop 2006). This choice is motivated by the fact that FFNNs include the possibility to reduce the dimensionality of the set of summary statistics via internal projections on lower dimensional subspaces (Bishop, 2006).
 
In FFNN regression models, linear combinations of the inputs --- i.e. the summary statistics --- are transformed to compute the values $z_j$, $j=1,\dots,H$ at the $H$ {\it hidden units} 
\begin{equation}
\label{eqn:ffnn1}
z_j=h(\sum_{k=1}^D w_{jk}^{(1)} {\bf s}^k +  w_{j0}^{(1)}),\quad j=1,\dots,H,
\end{equation}
where ${\bf s}^k$, $k=1, \dots , D$, is the $k^{th}$ component of ${\bf s}$, $H$ is the number of hidden units in the network, $D$ is the dimension of the vector ${\bf s}$ of summary statistics, the $w_{jk}^{(1)}$'s, $k=0, \dots , D$, $j=1, \dots , H$ are the weights of the first layer of the  neural network, and $h$ denotes the logistic function. Because the number of hidden units $H$ is typically smaller that the number of summary statistics $D$, equation (\ref{eqn:ffnn1}) reduces the initial dimension of the vector of summary statistics. The values $z_i$, $i=1,\dots,H$, of the hidden units are then linearly combined to produce the output $g_{\bf w}({\bf s})$ of the FFNN
\begin{equation}
\label{eqn:ffnn2}
g_{\bf w}({\bf s})=\sum_{j=1}^H (w_{j}^{(2)} z_z +w_{0}^{(2)}),
\end{equation}
where the $w_{j}^{(2)}$'s, $j=0, \dots ,H$, are the weights of the second layer of the neural network. Note that logistic regression can easily be performed within the FFNN framework by transforming the linear combination of equation (\ref{eqn:ffnn2}) with a logistic function. The extension to general discrete parameters is obtained using a softmax transformation (see Bishop 2006).

We use FFNNs for fitting both $m(\textrm{\bf s}_i)$ and $\log \sigma^2(\textrm{\bf s}_i)$ (Nix and Weigend 1995). The weights ${\bf w}$ of a first FFNN are found by minimizing the regularized least-squares criterion 
$$
\sum_{i=1}^n\{\phi_i-g_{\bf w}({\bf s})\}^2 K_{\delta}(\|\textrm{\bf s}_i - \textrm{\bf s}\|) + \lambda \| {\bf w} \|^2,
$$
where $\lambda$ represents the regularization parameter, called weight-decay parameter in the neural network literature. The weights of a second FFNN estimating the conditional variance, are found by minimizing
$$
\sum_{i=1}^n\{\log((\phi_i - \hat{m}(\textrm{\bf s}_i ))^2)-g_{\bf w'}({\bf s})\}^2 K_{\delta}(\|\textrm{\bf s}_i - \textrm{\bf s}\|) + \lambda \| {\bf w'} \|^2.
$$

Similarly to equation (\ref{eqn:adj_locl}), parameter adjustment under the NCH model can be performed as follows
\begin{equation}
\label{eqn:adj_nch}
\phi_i^* =  \hat{m}(\textrm{\bf s}) + \left( \phi_i  - \hat{m}(\textrm{\bf s}_i ) \right) \times \frac{\hat{\sigma}(\textrm{\bf s})}{\hat{\sigma}(\textrm{\bf s}_i )}   , \quad i = 1, \dots, M.
\end{equation}
Assuming that $\phi_i=m({\bf s}_i)+\sigma({\bf s}_i)\zeta_i$ corresponds to the true relationship between $\phi_i$ and ${\bf s}_i$, then $(\phi_i^*)$ forms a random sample from the distribution $p(\phi|{\bf s})$ provided that $\hat{m}$ could be considered equal to $m$ and $\hat{\sigma}$ equal to $\sigma$.

Similarly to the LocL ABC method, a tolerance error, $\delta$, is  allowed, and the adjusted parameters, $\phi_i^*$, are weighted by $K_{\delta}(\| \textrm{\bf s}_i - \textrm{\bf s} \|)$. Furthermore, to warrant that the adjusted parameters, $\phi_i^*$, obtained from equation (\ref{eqn:adj_locl}) or (\ref{eqn:adj_nch}), fall in the support of the prior distribution, we sometimes consider transformations of the original responses. Parameters that lie in an interval are transformed via the logit function, and nonnegative parameters are transformed using a logarithm. These transformations have the further potential advantage of stabilizing the variance of the response when performing regression (Box and Cox 1964).

\paragraph{Iterated importance sampling.}
A second change to the LocL ABC algorithm converts the single-stage regression based ABC method into a multi-stage algorithm in which estimations are improved iteratively (Liu 2001; Sisson et al. 2007). In practice, we implemented a two-stage algorithm. 
The logic of using a two-stage algorithm is that the second run can control a first run with a high acceptance rate,  and adaptively builds a better approximation of the posterior distribution.  If the two empirical distributions obtained after each step of the ANCH algorithm agree, then the results can be pooled to form a larger approximate sample from the posterior distribution.
 
The adaptive NCH algorithm can be described as follows: Starting from a sample $(\phi^1_i)$ obtained from a first NCH ABC run, the adaptive step of the algorithm consists of estimating the support $\Delta_1$ of the sample. Then new parameters are proposed from the conditional prior distribution $p_{\Delta_1}$ given that they fall in $\Delta_1$. This can be implemented using a simple rejection step. For $\phi$ having a moderate number of dimension, this is usually achieved at a computational cost which can be considered significantly lower than the cost of simulating from the generating distribution, $p({\bf s}|\phi)$. Using this new set of parameters, a second sample, $(\phi^2_i)$, can be formed using the NCH method again. To compensate for the fact that we do not sample from the prior distribution, each value $\phi^2_i$ should in principle be weighted by $p(\phi^2_i)/p_{\Delta_1}(\phi^2_i)$. Here, the importance weights need not to be computed because we have $p_{\Delta_1}(\phi^2_i) = p(\phi^2_i)/p(\Delta_1)$, which means that the ratio $p(\phi^2_i)/ p_{\Delta_1}(\phi^2_i)$ does not depend on $i$. In this two-stages approach, we suppose that $\Delta_1$ approximates the support of $p({\bf s}|\phi)$ accurately. For multi-dimensional parameters $\phi$, we estimate the support of the distribution using support vector machines (SVM, Sch\"olkopf et al. 2001).

\section{Examples of implicit statistical models}

In this section, we present three examples of implicit statistical models, two of which have received considerable attention in population genetics, and the last one has served to illustrate  {\it indirect inference} (Gourieroux et al. 1993; Heggland and Frigessi 2004). Using these examples, we performed an empirical evaluation of the relative performance of three regression-based approximate Bayesian algorithms, the local linear regression model (LocL ABC model), the non-linear conditional heteroscedastic model (NCH model), and its adaptive implementation (ANCH model). We used the R programming language to implement the LocL, NCH and ANCH algorithms. Least squares adjustment for neural networks was implemented using the R package {\tt nnet} (R Core Team 2008). Model choice for neural networks was based on a Bayesian (or regularization) approach (Ripley 1996). We used $H=4$ hidden units and the weight-decay regularization parameter was set equal to $\lambda = 0.001$.  Increasing the weight-decay parameter will increase the bias of the estimator, but it will also decrease its variance. At this stage, cross-validation might be a useful alternative approach to the choice of a FFNN model, but the previous values for $\lambda$ and $H$ proved to work well in the examples considered here. In the ANCH algorithm, the support of the conditional density was estimated using a SVM $\nu$-regression algorithm ($\nu = 0.005$) as implemented in the R package {\tt e1071} based on the public library {\tt libsvm} (Chang and Lin 2001).

\paragraph{Two examples in population genetics.} There has been tremendous interest in simulation-based inference methods in evolutionary biology during the last decade (Fu and Li 1997; Pritchard et al. 1999; Fagundes et al. 2007). In these applications, the inference of demographic and genetic parameters depends on the so-called coalescent approximation which describes, in a probabilistic fashion, the ancestry of genes represented in a sample. Coalescent models provide good examples of implicit statistical models for which a straightforward stochastic generating mechanism exists, but the likelihood is usually computationally intractable (see the Supplementary Material for further information).

\paragraph{Example 1.} Given a set of $n$ DNA sequences, the first problem concerns the estimation of the effective mutation rate, $\theta > 0$, under the infinitely-many-sites model.  In this model, mutations occur at rate $\theta$ at DNA sites that have not been hit by mutation before. If a site is affected by mutation, it is said to be {\it segregating} in the sample. In this example, the summary statistic, {\bf s}, is computed as the number of segregating sites. Note that {\bf s} is not a sufficient statistic with respect to $\theta$ (Fu and Li 1993). The generating mechanism for {\bf s} can be described as follows. 
\begin{enumerate}
\item Simulate $L_n$, the length of the genealogical tree of the $n$ sequences, as the sum of independent exponential random variables of rate $(j-1)/2$, $j = 2, \dots, n$. 
\item Generate {\bf s} according to a Poisson distribution of mean $\theta L_n/2$. 
\end{enumerate}
The simulation of $L_n$ can be derived from the formula $L_n=\sum_{j=2}^n j Y_j$ that gives the total length of the tree as a function of the inter-coalescence times denoted by $Y_j$, $j=2,\dots , n$. The inter-coalescence times correspond to the times during which the sample has $j$ ancestors, $j=2,\dots,n$ (Tavar\'e et al. 1997). In a wide range of models in population genetics, the inter-coalescence times form independent random variables distributed according to the exponential distributions of rate $j(j-1)/2$, $j=2,\dots,n$. A more detailed description of the coalescent process can be found in (Tavar\'e 2004).

We computed the posterior distribution of the effective mutation rate $\theta$ given the observation of ${\bf s} = 10$ segregating sites in a sample of $n = 100$ DNA sequences. The prior distribution for the parameter $\theta$ was taken to be the exponential distribution of mean 50, which was meant to represent a vague prior. An sample from the posterior distribution was obtained using a direct rejection algorithm accepting only parameters that produced 10 segregating sites exactly. Ten millions of replicates were generated resulting in a sample of size $39,059$ after rejection.

For the ABC algorithms, we performed inference of the posterior distribution using a total of 2,000 simulations of the bivariate vector $(\theta,{\bf s})$. We recorded the three quartiles and the $0.025$ and $0.975$ quantiles of the approximate posterior distributions computed by the three algorithms, and we compared these 5 quantiles $Q_k$, k=1,\dots,5, with the corresponding empirical quantiles, $Q_k^0$, obtained from the exact sample. For values of the tolerance rate between 0 and 1 and for each quantile, $Q_k$, the accuracy of each algorithm was assessed by the relative median absolute error  defined as the median of  $(Q_k - Q_k^0)/Q_k^0$
 computed over 150 runs. In addition, we measured the discrepancy between each approximate distribution and the empirical posterior distribution using the sum of the RMAE's over all quantiles.  In the ANCH algorithm, the support was estimated as the range  of the empirical distribution -- i.e., the $(0, \max)$ interval --  obtained after a first run. A total of 1,000 replicates were used in the first run and the tolerance rate, denotes as $P_{\delta}$, was set to $75\%$.

\begin{figure}[h]
\begin{center}
\includegraphics[height = 6.5cm]{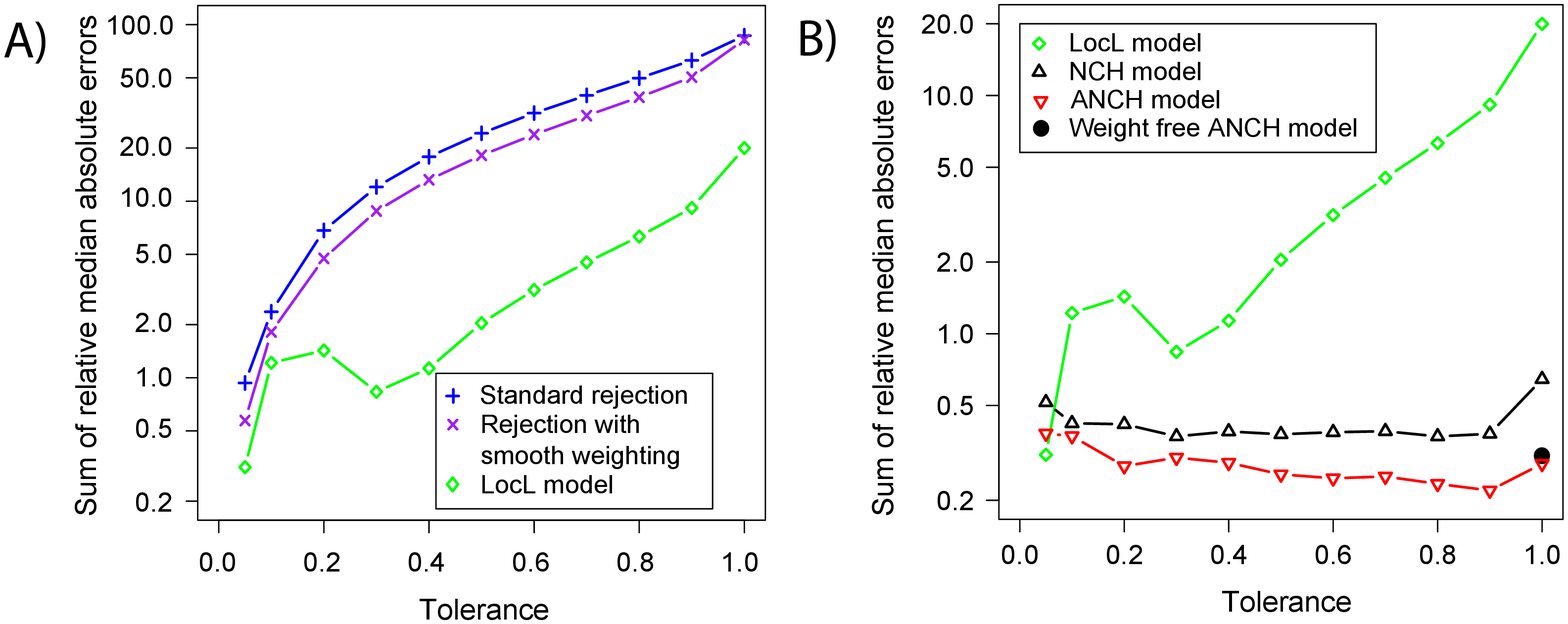}
\end{center}
\caption{\bf }
\label{fig:RMAE}
\end{figure}

Comparisons with standard rejection algorithms were first conducted. We found that the posterior distribution obtained from the rejection methods deviated from the empirical posterior distribution significantly for tolerance rates larger than 10\% (Figure \ref{fig:RMAE}A). The LocL model approximated the posterior distribution accurately for small tolerance rates ($\leq 20\%$), but the performances of the LocL method deteriorated as the tolerance rate increased (Figure \ref{fig:RMAE},  curves with diamonds). The performances of the NCH model were significantly less sensitive to the tolerance rate, staying at values close to the optimum achieved by the LocL model (Figure \ref{fig:RMAE}B,  curve with triangles point-up). The adaptive NCH algorithm achieved even superior performances for values of the tolerance rate ranging between 0 and $90\%$ (Figure \ref{fig:RMAE}B,  curve with triangles point-down). The black dot in Figure \ref{fig:RMAE}B represents the performance of the ANCH algorithm without weighting and allowing total acceptance. Having eliminated the concept of rejection in the approximate Bayesian algorithm, i.e. setting $P_{\delta} = 1$, the accuracy of the algorithm remained close to the optimum achieved by all algorithms. This first example illustrates the benefit of the NCH model over the LocL model and the other rejection methods. The additional gain of the adaptive step stems from the use of a vague prior, which, in this case, gave low weight to the region of posterior values. We note, however, that for the smallest tolerance rate, the RMAE is smaller in the LocL model than in the NCH models. Because of the additional level of complexity introduced in FFNNs, FFNNs require more data points to be trained than local linear regression, but this not an issue for higher tolerance rates.

\paragraph{Example 2.} Turning to a more complex problem in which the posterior distribution could not be estimated easily, we considered an exponentially growing population model with 3 parameters. Similarly to Weiss and von Haeseler (1998), a population of size $N_A$ started to grow exponentially $t_0$ years ago to reach a present size of $N$ individuals where $N=N_A/\alpha$ for a value $\alpha \in (0,1)$. We performed inference on the two parameters $N_A$, $t_0$ and $\alpha$ was considered as a nuisance parameter. In this example the data consisted of a sample of $n$ individuals genotyped at a multilocus subset of independent microsatellite markers (see e.g. Zhivotovsky et al. 2003). Microsatellite loci are characterized by a motif of two to four nucleotides that may repeat itself several times, and the data are recorded as number of repeats for each individual. 

The generating mechanism for the implicit model can be described as follows.
\begin{enumerate}  
\item Simulate candidate coalescent genealogies with $n$ tips in a growing population for each marker, 
\item Superimpose mutations on the tree branches according to a specific mutation model. 
\end{enumerate}
Step 1 requires simulating coalescence times in a coalescent model with varying population size (see Tavar\'e 2004). In step 2, we used the single-step mutation model, that can be viewed as a simple random walk for which the +1 and -1 steps are equally likely (Ohta and Kimura 1973).

To capture the pattern of genetic variation, we computed six summary statistics previously reported to be sensitive to the genetic diversity in the sample and to the intensity of the demographic expansion. The amount of genetic diversity was measured by the mean (over the loci) of the variance in the number of repeats and by the mean of their heterozygosities (Pritchard and Feldmann 1996). For the demographic pattern, we used two imbalance indices studied by King et al. (2000), the interlocus statistic introduced by Reich and Goldstein (1998), and the expansion index of Zhivotovsky et al. (2000).
We also computed a seventh summary statistic based on an observation of Shriver et al. (1997) who studied the distribution, $S_K$, of pairwise comparisons that differ by $K$ repeat units. This distribution has its peak at the value $0$ for a recent expansion, and the peak shifts to the value $1$ for older expansions. To compute the seventh statistic, we averaged the quantity $S_1-S_0$ over all the loci.

We took a uniform prior distribution ranging from 0 to 100,000 years for the onset of the expansion, a uniform distribution over the interval (0, 10,000) for the ancestral population size, and a uniform distribution over the interval $(1,6)$ for $-\log_{10}(\alpha)$. 

One hundred test data sets were generated using $t_0 = 18,000$ years for the date of onset of the expansion, $N_A = 1,500$ for the ancestral population size, and $\alpha = 0.0012$ ($\log_{10}(\alpha) = -2.92$) for the ratio of the ancestral size to the present population size. These values were very similar to those used in Pritchard et al. (1999) in a study of the Y chromosome in humans.  For each algorithm, we computed the posterior distribution of the 3 parameters given the observation of the 7 summary statistics in a sample of $n = 100$ individuals surveyed at 50 microsatellite loci.

For the three algorithms, we generated samples from the posterior distribution using 2,000 replicates from the implicit model. We recorded the quartiles and the $0.025$ and $0.975$ quantiles of the output distributions for the  three ABC algorithms, and we compared these 5 values for the conditional distributions of each parameter $t_0$, $N_A$, and $\alpha$. The median value of each quantile was then computed over 100 runs.

\begin{figure}[h]
\begin{center}
\includegraphics[height = 18cm]{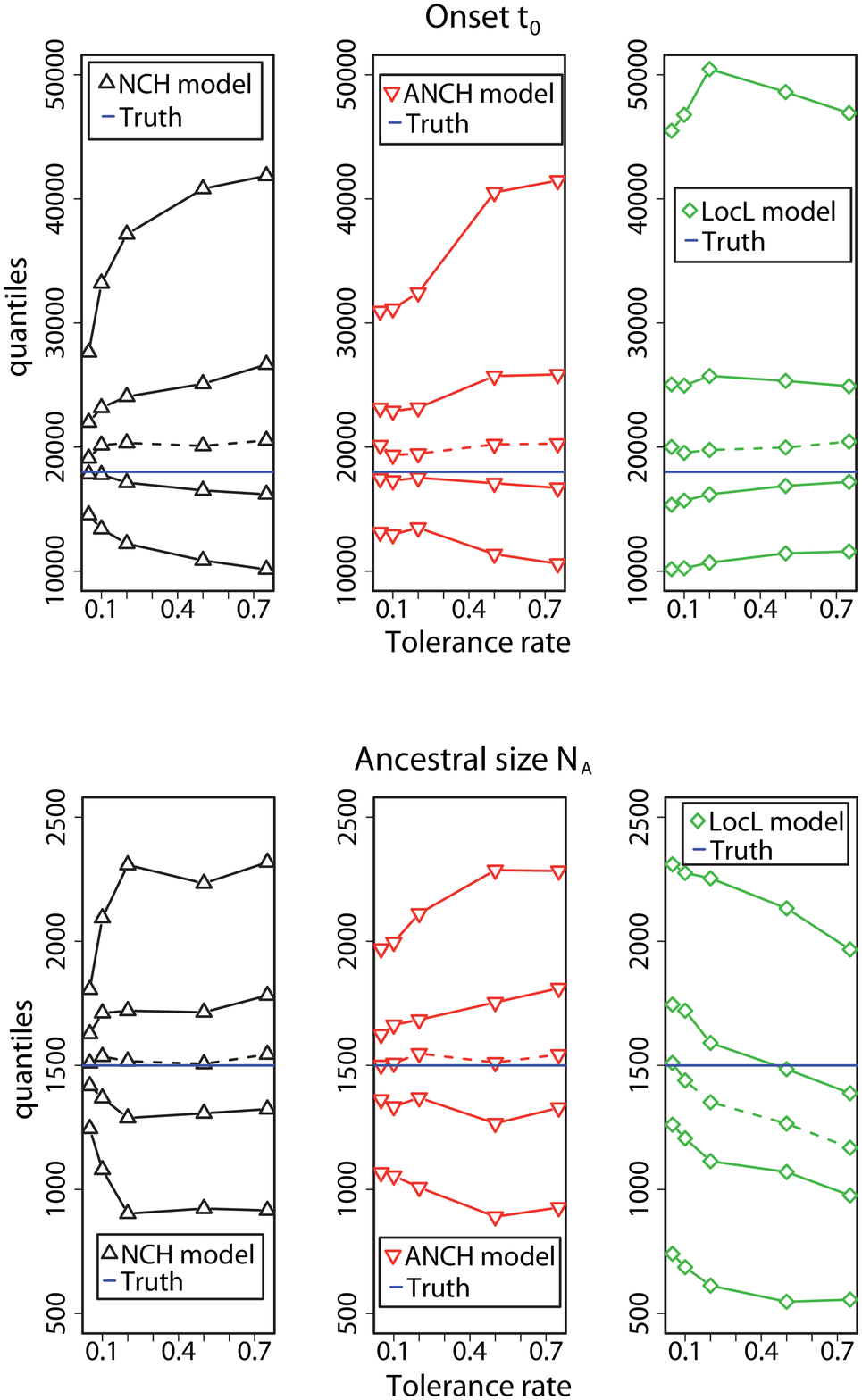}
\end{center}
\caption{\bf }
\label{fig:all_quantiles}
\end{figure}

For each of the three parameters, the median estimates of the quantiles of the marginal posterior distribution were very similar in the NCH model and in the ANCH implementation (Figure \ref{fig:all_quantiles}, black and red curves). For large tolerance rates ($P_{\delta} \geq 50\%$), we observed a strong agreement with the values estimated by the LocL model (Figure \ref{fig:all_quantiles}, green curves) used with small tolerance rates ($P_{\delta} = 5\%$), indicating that the NCH model can efficiently exploit simulation results that fall far apart from the observed values of the summary statistics. The performances of the LocL model decreased as the tolerance rate increased above $20\%$, and the estimation of the conditional distribution of the ancestral size provided evidence that the bias was substantial (Figure \ref{fig:all_quantiles}, bottom right panel). The three algorithms gave similar results regarding the estimation of $\alpha$ (results not shown).

To further compare the performances of the NCH model and its ANCH variant, we studied a particular simulated data set  corresponding to the same ground truth as before. After running the three algorithms 100 times for each tolerance rate, the variance of each quantile in the posterior distribution was of order three times higher in the NCH model than in the ANCH algorithm (2,000 simulations for each algorithm, Table 1). Clear evidence of the stabilization phenomenon was also obtained under the infinitely-many-sites model. Given ${\bf s}=10$ segregating sites, we ran the ANCH algorithm  using 200 simulations of the implicit model at each step (tolerance rate $P_\delta = 85\%$). To compare estimations obtained after the initial step with those obtained after adapting the support, we replicated the estimation procedure 100 times. The reduction in variance ranged from a factor 2.7 to a factor 34.7. The highest reduction in variance was obtained for the estimation of the upper quantile.

\paragraph{Example 3.} Our third example arose from a totally different context, and was formerly studied by Heggland and Frigessi (2004) using indirect inference. The connection of indirect inference to ABC is the following.  Indirect inference (Gourieroux et al. 1993) is a non-Bayesian method that proceeds with 3 steps. 1) an auxiliary model is introduced, usually as a simplified version of the true model. 2) Estimates of the parameters in the auxiliary model are obtained and play the role of summary statistics. These estimates can be obtained, for example, by maximizing the likelihood in the auxiliary model. 3) An estimate of the parameter $\phi$ is built by minimizing a weighted Euclidean distance between simulated summary statistics and the observed summary statistics. Note that the introduction of an auxiliary model can also be a useful means of finding informative summary statistics for ABC methods.

The model considered in (Heggland and Frigessi 2004) was a queueing system with a first-come-first-serve single-server queue (G/G/1). The service times were uniformly distributed in the interval $[\theta_1,\theta_2]$, and the inter-arrival times had exponential distribution with rate $\theta_3$. Let $W_n$ be the inter-arrival time of the $n$th customer and $U_n$ be the corresponding service time. The process of inter-departure times $\{Y_n,\, n=1,2,\dots\}$ can be described by the following generative algorithm
$$
Y_n=
\left\{
\begin{array}{ll}
U_n \, ,  & \mbox{if } \sum_{i=1}^n W_i \leq \sum_{i=1}^{n-1} Y_i, \\
 & \\
U_n + \sum_{i=1}^n W_i - \sum_{i=1}^{n-1} Y_i \, , & \mbox{otherwise.}
\end{array}
\right.
$$
Bayesian inference on $(\theta_1,\theta_2,\theta_3)$ was done by assuming that only the inter-departure times were observed. Because the inter-arrival times were unobserved, likelihood-based inference would involve high-dimensional integration. 

We generated a test data set with $n=50$ successive inter-departure time observation using $\theta_1=1$, $\theta_2=5$, and $\theta_3=0.2$. We set a uniform prior over $[0,10]$ for $\theta_1$, $\theta_2-\theta_1$, and for $\theta_3$. To investigate the sensitivity of the NCH model to the number of summary statistics, we ran the ABC-NCH algorithm using 5, 10 and 20 summary statistics. Here, the set of summary statistics included the minimum and the maximum of the inter-departure times and the 3, 8 and 18 equidistant quantiles of the inter-departure times. We used 10,000 replicates and the tolerance rate was set to the value $P_{\delta} = 50\%$ resulting in a posterior sample of size 5,000. 

\begin{figure}[h]
\begin{center}
\includegraphics[height = 14cm]{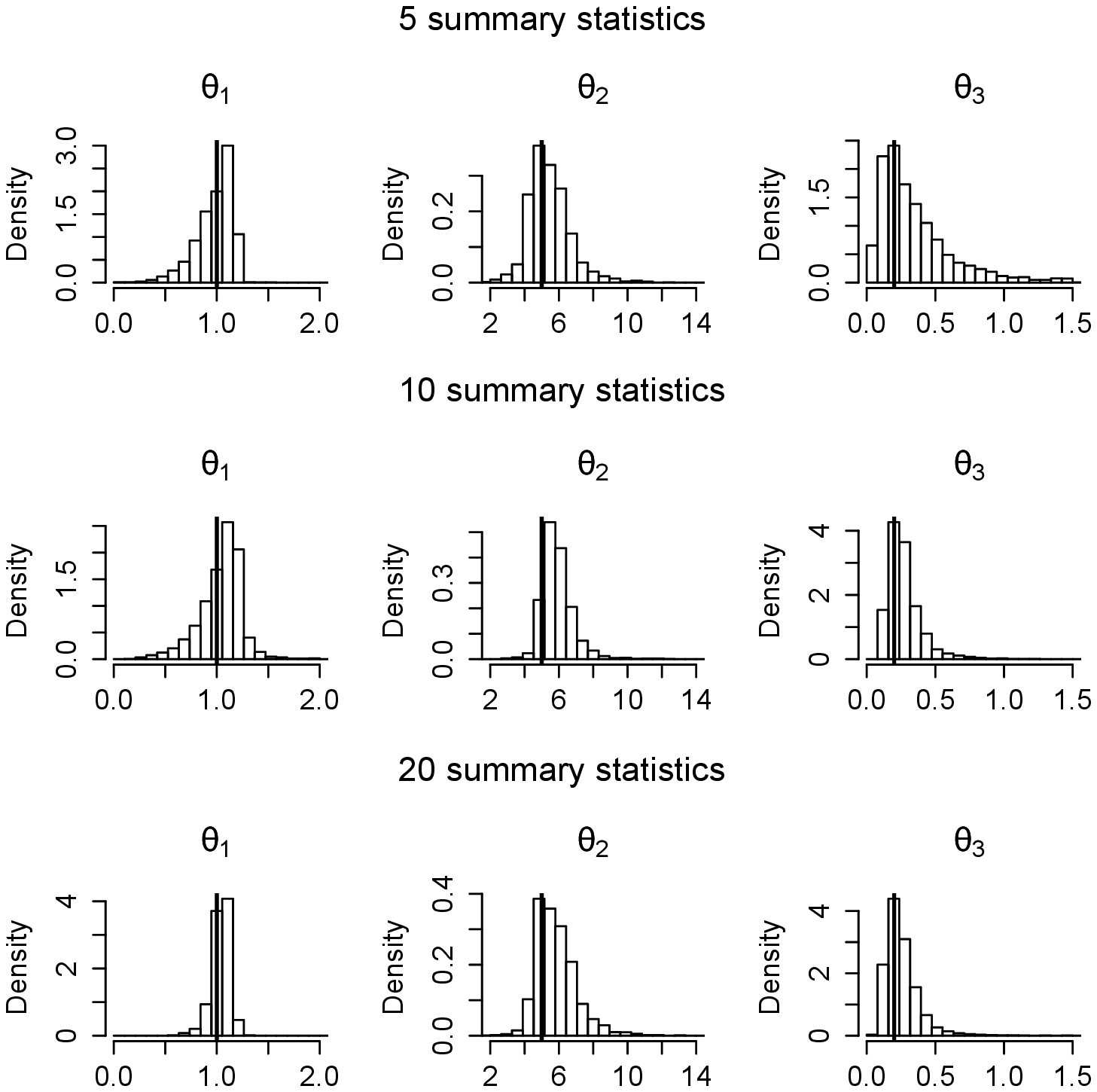}
\end{center}
\caption{\bf }
\label{fig:9hist}
\end{figure}

Figure \ref{fig:9hist} shows that posterior distributions of the 3 parameters had their mode and median values close to the ground truth values whatever the number of summary statistics. This provided evidence that the NCH ABC algorithm was robust to an increase in the dimensionality of the set of summary statistics. In addition, we found that the posterior distributions were more concentrated around the true values when 20 summary statistics were used. 

To investigate the variability of the ABC algorithms from one run to the other, we ran the LocL, NCH and ANCH algorithms 100 times on the same data set. We used 2,000 replicates (a rather small number) in order to observed an exaggerated variability, and we varied the tolerance rate from 0 to 1. 

\begin{figure}[h]
\begin{center}
\includegraphics[height = 12cm]{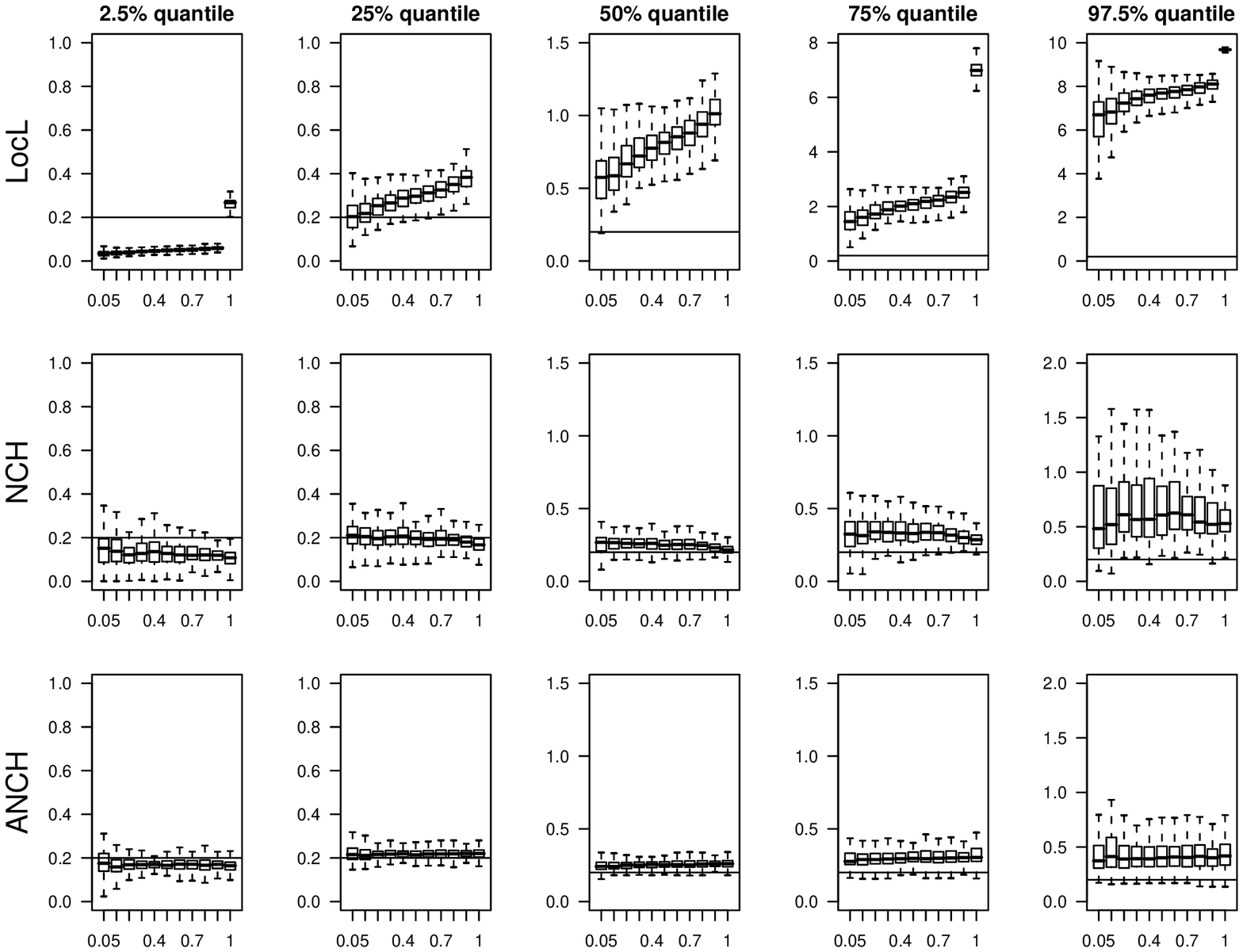}
\end{center}
\caption{\bf }
\label{fig:param3}
\end{figure}

Figure \ref{fig:param3} displays estimated posterior quantiles for $\theta_3$.
The LocL ABC algorithm was the less variable algorithm, but the posterior credibility intervals produced by this method were wider than those produced by the NCH and ANCH algorithms. The latter were less sensitive to the variation of the tolerance rates, and the medians of the posterior distributions were closer to the true value in the non-linear models than in the LocL model. Compared to the NCH algorithm, the ANCH algorithm reduced the variance of the quantile estimates. We suspect that the variance of the posterior quantile estimates came from the use of local optimization during the learning phase of the feed-forward neural networks. For $\theta_1$ and $\theta_2$, the LocL ABC and the NCH ABC algorithms led to similar approximate posterior distributions when the tolerance rate was set at values close to zero in the LocL ABC algorithm.

\section{Discussion}

Approximate Bayesian computation encompasses a wide range of useful methods for making inference in implicit statistical models. In this context rejection algorithms have greatly benefited from ideas coming from regression-based conditional density estimation. So far conditional density estimation in ABC approaches has relied on linear adjustment exclusively (Beaumont et al. 2002).  While the linear regression-based ABC method can approximate posterior distribution accurately, this is usually achieved at the expense of a heavy computational load. For example, using the LocL ABC method for estimating parameter in complex models of modern human expansion, Fagundes et al. (2007) required an amount of computational time equivalent to 10 CPU-months. To increase the tolerance of the algorithm, we have proposed to use non-linear regression-based ABC. In three examples, non-linear neural networks proved to be able to reduce computational generation costs significantly.

A heuristic reason why neural networks worked well when the number of summary statistics was large is that their first layer allows a nonlinear projection onto a subspace of much lower dimensionality, and nonlinear regression can then be performed using the reduced number of projection variables. Increasing the number of summary statistics has a dramatic effect on the variability of the estimators of the conditional mean $\hat{m}(\textrm{\bf s}_i )$ and variance $\hat{\sigma}^2(\textrm{\bf s}_i )$ and consequently inflates the variances of the estimated posterior distributions. The variance can be reduced with the Bayesian predictive approach of Ripley (1996) that consists of training a large number of FFNNs for each conditional regression and averaging the results over the replicate networks. In addition Bayesian neural network theory includes general rules for choosing appropriate regularization parameters which makes the method rather automatic. Compared to other regression models, neural networks share many properties of projection pursuit regression (Friedman and Stuelze 1981), which may then lead to equivalent performances. As well SVM have gained increased popularity in machine learning approaches during the recent years (Vapnik 1998), and the algorithms described here could be modified to include SVM regression without change in spirit.

A second justification for using feed-forward neural networks is their ability to implement probabilistic outputs, hence allowing a unified Bayesian treatment of model choice. Indeed model choice may be performed by considering the model itself as an additional parameter to infer. Beaumont (2007) proposed to estimate the posterior probability of each candidate model by an approach based on a weighted multinomial logistic regression procedure. This approach is an extension of logistic regression to more than two categories, and it is equivalent to the use of a multinomial log-linear model. As they pertain to a more flexible class of models, neural networks may achieve equal or better predictive values than multinomial logistic regression (Ripley 1996).

The ABC approach has been recently used in connected domains like human population genetics (Pritchard et al. 1999, Fagundes et al. 2007), epidemiology (Tanaka et al. 2006, Toni et al. 2009, Blum and Tran 2008) or for the evolution of protein networks (Ratmann et al. 2007). It has also recently been suggested in the context of compositional data (Butler and Glasbey 2008), and applied in extreme value theory (Bortot et al. 2007) and Gibbs random fields (Grelaud et al. 2008). Although inference from synthetic data that mimic observations has a long lasting record in frequentist statistics (Diggle and Gratton 1984, Gourieroux et al. 1993), ABC is still in its infancy. Because the ABC method combines the power of simulating from stochastic individual-based models with sound methodological grounds from Bayesian theory, it has the potential to open doors to inference in many complex models in ecology, evolution, and epidemiology, or other domains like the social science. Improved statistical ABC models, like those presented in this study, will then be useful to deal with increased model complexity, and with the need to raise the dimension of the vector of summary statistics. An R code for performing ABC with the NCH model is available at the author's website.

\vspace{1cm}

\paragraph{Acknowledgements}
The authors acknowledge the support of the ANR grant MAEV and M. B. acknowledges the support of the Rh\^one-Alpes Institute of Complex Systems (IXXI). We are also grateful to Oscar Gaggiotti and two anonymous reviewers for their helpful comments.

\clearpage
\newpage


\clearpage
\newpage

\clearpage
\newpage


\begin{center} {\sc Figure legends} \end{center}

%
%
%

\noindent {\bf Figure 1. \quad} Relative median absolute error (RMAE) when estimating the quantiles of the posterior distribution of $\theta$ as a function of the tolerance rate.
In panel A), the errors obtained with the regression algorithms are compared to the errors obtained with the local linear adjustment. In panel B), the different regression methods are compared. For the quartiles and for the 0.025 and 0.975 quantiles, relative errors between approximate 
quantiles computed by the ABC methods and empirical values from the posterior distribution
were computed over the 150 replicates. The sum of the RMAE's was obtained by summing, over the 5 quantiles, the values of the relative median absolute differences. The black dot corresponds to the rejection-free ANCH algorithm (tolerance rate $P_{\delta}= 100\%$) without including the Epanechnikov weights.

\vspace{1cm}

\noindent {\bf Figure 2. \quad} The posterior quantiles of the date of onset of expansion and the ancestral population size for the NCH and the ANCH methods (left) and LocL ABC method (right).  The quantiles are plotted against the tolerance rate for the NCH and the ANCH methods (left) and LocL ABC method (right). For each algorithm, the curves represent the 0.025, 0.25, 0.5, 0.75, and 0.975 quantiles of the posterior distribution. These values correspond to the median over the 100 test data sets. The ground truth values are represented as the blue lines. The values of the quantiles estimated by the NCH and ANCH methods using a tolerance rate $P_{\delta}=75\%$ match with those obtained from the LocL algorithm using $P_{\delta}=5\%$.

\vspace{1cm}

\noindent {\bf Figure 3. \quad} The posterior quantiles of $\theta_1$, $\theta_2$, and $\theta_3$, using the NCH ABC method, with 5, 10, and 20 summary statistics. The vertical lines correspond to the true values of the parameters that were used when simulating the data set. The tolerance rate was set at $50\%$ and a total of 10,000 simulations were performed.

\vspace{1cm}

\noindent {\bf Figure 4. \quad} The boxplot of the posterior quantiles for $\theta_3$ as a function of the tolerance rate. The different ABC methods were run 100 times each using 2,000 simulations at each run.

\clearpage
\newpage

\pagestyle{empty}

%
%

\clearpage
\newpage

\clearpage
\newpage

\clearpage
\newpage

\clearpage
\newpage

\begin{center} {\sc Table caption} \end{center}

\noindent {\bf Table 1. \quad} The benefit of adapting the support. Infinitely-many-sites: Ratios of variances for the quantiles estimated after the first NCH step and the second NCH step of
the ANCH algorithm (acceptance rate $P_{\delta} = 85\%$ in each step). Expansion
model: Ratios of variances for the quantiles estimated by the NCH  and the ANCH
algorithm (2,000 simulations in each algorithm, $P_{\delta} = 75\%$). P-values
were computed according to the F-test.
\vspace{1cm}

\clearpage
\newpage

\begin{table}
\begin{tabular}{l|ccccc}
~ & \multicolumn{5}{c}{Posterior distribution quantiles}\\
& & & & &\\
Model parameters      & 2.5$\%$ &       25$\%$ &        50\%   &     75\% &     97.5\% \\ 
\hline
& & & & &\\
{\sc Infinitely-many-sites}  & & & & &\\
& & & & &\\
Mutation rate $\theta$ & & & & &\\
Var. ratio & 2.75 & 3.16 & 3.37  & 5.46 & 34.76\\
$P$-values &
1.27e-06 & 1.67e-08 &1.76e-09 & 6.88e-15 & 0 \\ 
& & & & &\\
\hline
& & & & &\\
{\sc Expansion model} & & & & &\\
& & & & &\\
Onset $t_0 $& & & & &\\
Var. ratio & 1.84 & 2.95 & 3.69  & 3.53 & 2.97\\
$P$-values &
1.28e-03 & 8.12e-08 & 1.74e-10 & 6.16e-10 & 6.12e-08 \\
& & & & &\\
Ancestral pop. size $N_A$ & & & & &\\
Var. ratio & 1.65 & 3.23 & 6.83  & 4.58  & 3.20 \\
$P$-values &
6.20e-03 & 7.15e-09 & 0 & 2.3e-13 & 9.34e-09 \\ 
& & & & &\\
Ratio of pop. sizes $\alpha$ & & & & &\\
Var. ratio & 0.09 & 0.87 & 3.83  & 1.97 & 7.66 \\
$P$-values &
1 & 0.75 & 5.89e-11 & 4.24e-04 & 0 \\
& & & & &\\
\hline
\end{tabular}
\caption{\bf }
\end{table}

\clearpage
\newpage


\end{document}